\documentclass[publish]{smj}

\usepackage{enumitem}
\usepackage{amssymb}
\usepackage{amsmath} 
\usepackage{amsthm}
\usepackage{subcaption}
\usepackage{adjustbox}

\newtheorem{Result}{Result}
\newtheorem*{Result5b}{Result 5b}

\Author{Alan Huang\Affil{1}, 
}
\AuthorRunning{Alan Huang}

\Affiliations{

\item School of Mathematics and Physics,
	University of Queensland, St Lucia, QLD, AUS

}   

\CorrAddress{Alan Huang, 
             School of Mathematics and Physics,
             University of Queensland, 
             St Lucia, 
             QLD, 4066,
             Australia}
\CorrEmail{alan.huang@uq.edu.au}
\CorrPhone{(+617)\;3365\;2315}
\CorrFax{(+617)\;3365\;3328}

\Title{Mean-parametrized Conway-Maxwell-Poisson regression models for dispersed counts}
\TitleRunning{Mean-parametrized Conway-Maxwell-Poisson models}

\Abstract{
Conway-Maxwell-Poisson (CMP) distributions are flexible generalizations of the Poisson distribution for modelling overdispersed or underdispersed counts. The main hindrance to their wider use in practice seems to be the inability to directly model the mean of counts, making them not compatible with nor comparable to competing count regression models, such as the log-linear Poisson, negative-binomial or generalized Poisson regression models. This note illustrates how CMP distributions can be parametrized via the mean, so that simpler and more easily-interpretable mean-models can be used, such as a log-linear model. Other link functions are also available, of course. In addition to establishing attractive theoretical and asymptotic properties of the proposed model, its good finite-sample performance is exhibited through various examples and a simulation study based on real datasets. Moreover, the MATLAB routine to fit the model to data is demonstrated to be up to an order of magnitude faster than the current software to fit standard CMP models, and over two orders of magnitude faster than the recently proposed hyper-Poisson model.
}

\Keywords{
Conway-Maxwell-Poisson distribution; Count data; Generalized linear model; Overdispersion; Underdispersion
}

\begin{document}

\maketitle

\vspace{-5mm}
\section{Introduction}
\label{se:introduction}
\vspace{-5mm}

Conway-Maxwell-Poisson (CMP) distributions have seen a recent resurgence in popularity for the analysis of dispersed counts \citep[e.g.][]{SMKBB2005, SS2010, LGG2008, LGG2010}. Key features of CMP distributions include the ability to handle both overdispersion and underdispersion, containing the classical Poisson distribution as a special case, and being a continuous bridge between other classical distributions such as the geometric and Bernoulli distributions. CMP distributions are also full probability models, making them particularly useful for predictions and estimation of event probabilities. 

Originally developed as a model for queueing systems with dependent service times \citep{CM1962}, CMP distributions have subsequently found novel applications in the modelling of clothing sales, number of syllables of words in the Hungarian dictionary \citep{SMKBB2005}, and in the regression modelling of airfreight breakages \citep{SS2010}, overdispersed counts of motor vehicle crashes \citep{LGG2008}, and underdispersed counts of motor vehicle crashes \citep{LGG2010}. See \citet{SMKBB2005} for a more detailed overview of the history, features and applications of CMP distributions.

One of the major limitations of CMP distributions is that it is not directly parametrized via the mean. Instead, the two model parameters, namely, the rate $\lambda \ge 0$ and the dispersion  $\nu \ge 0$, are interpreted through ratios of successive probabilities via
$$
\frac{P(Y=y-1)}{P(Y=y)} = \frac{y^\nu}{\lambda} \ .
$$
The moments of CMP distributions satisfy recursive formulas,
$$
E(Y^{r+1}) = \begin{cases}
\lambda E(Y+1)^{1-\nu} ,  & r=0 \\
\lambda \frac{d}{d\lambda} E(Y^r) + E(Y) E(Y^r)  , &  r > 0 \ ,
\end{cases}
$$
but cannot be solved in closed form. For the first moment, an approximation can be obtained as
$$
E(Y) \approx \lambda^{1/\nu} - \frac{\nu - 1}{2 \nu}  \ ,
$$
which is particularly accurate for $\nu \le 1$ or $\lambda > 10^\nu$ \citep{SMKBB2005}. However, this approximation can be inaccurate outside those ranges. Interestingly, the recent resurgence of CMP distributions in practice has been for analysing underdispersed counts, but this corresponds to $\nu >1$ which is precisely the region where the approximation fails, unless the rate $\lambda > 10^\nu$ is also very large.

The inability to model the mean directly has arguably limited the use of CMP distributions in practice, especially in regression scenarios where an applied scientist might sacrifice any perceived gains in model flexibility for a simpler, more easily interpretable approach, such as a log-linear Poisson, Negative-Binomial, or generalized Poisson \citep[GP,][]{CF1992, Famoye1993} regression model. Other alternatives, such as the log-linear hyper-Poisson regression model \citep{SS2013}, which is a special case of the  exponentially-weighted Poisson model \citep{RB2004}, and the Gamma-count model \citep{ZRBSM2014}, have also been proposed recently.  Being able to parametrize CMP distributions via the mean will therefore be invaluable as it allows CMP models to be directly compatible and comparable to a host of competing count regression models.

Of the above-mentioned models, the only two that can handle both underdispersion and overdispersion and have achieved some popularity in practice seem to be the CMP and GP models. 
\citet{SS2010} argue that GP models are less attractive than CMP models because GP models achieve underdispersion via a somewhat unnatural truncation of the support of the distribution. In fact, the support and the parameter space are functionally dependent. This is particularly undesirable in regression settings, as the level of underdispersion places an artificial restriction on the allowable range of means, and vice versa. In contrast, the support of CMP distributions is always the non-negative integers $\mathbb{N} = \{0,1,2,\ldots \}$ for any finite value of the dispersion. 

In this paper, we show how the means of CMP distributions can be modelled directly. We argue that this will promote their wider use in practice, putting them on the same level of interpretability and parsimony as competing, but less flexible, count regression models whilst retaining all the desirable features of CMP models. The simulation and data analysis examples in this paper reinforce that the proposed framework is a simple yet flexible and robust way to model counts. Moreover, the MATLAB routine for fitting the proposed models is up to an order of magnitude faster than the state-of-the-art \texttt{cmp} function from the \texttt{COMPoissonReg} R-package \citep{SL2015} for fitting standard CMP models, and over two orders of magnitude faster than the \texttt{hp.fit} function for fitting hyper-Poisson models in R, available from the website \url{http://www4.ujaen.es/~ajsaez/hp.fit.r} that accompanies the paper by \citet{SS2013}. The MATLAB routine can be downloaded from the Online Supplement.

\section{Parametrizing CMP distributions via the mean}
\vspace{-5mm}

The CMP distribution with rate parameter $\lambda$ and dispersion parameter $\nu$ has probability mass function (pmf) given by
$$
P(Y=y \, |\, \lambda, \nu) = \frac{\lambda^y}{(y!)^\nu} \frac{1}{Z(\lambda, \nu)} \ , \ y=0,1,2,\ldots \ ,
$$
where $Z(\lambda, \nu) = \sum_{y=0}^\infty \lambda^y/(y!)^\nu$ is a normalizing constant. A detailed summary of the key properties of CMP distributions can be found in \citet{SMKBB2005} -- here, we focus on a novel reparametrization via the mean.

Note that for a single sample of data, any reparametrization of the CMP distribution is equivalent (although some parametrizations can be easier to estimate than others). However, for regression settings in which the count responses may change depending on a set of covariates, it is more convenient and interpretable to model the mean $E(Y)$ of the distribution directly. To this end, let $\mu = E(Y)$ be the mean of the distribution. The CMP distribution with mean $\mu \ge 0$ and dispersion $\nu \ge 0$ can then be characterized by the pmf,
\begin{equation}
\label{eq:compmupdf}
P(Y = y \, | \, \mu, \nu) =   \frac{\lambda(\mu, \nu)^y}{(y!)^\nu} \frac{1}{Z(\lambda(\mu, \nu), \nu)}, \ y=0,1,2,\ldots \ ,
\end{equation}
where the rate $\lambda(\mu, \nu)$ is a function of $\mu$ and $\nu$, given by the solution to 
\begin{equation}
\label{eq:lambdamu}
0 = \sum_{y=0}^\infty (y - \mu) \frac{\lambda^y}{(y!)^\nu} \ .
\end{equation}
We call the distribution defined by (\ref{eq:compmupdf})--(\ref{eq:lambdamu}) the CMP$_\mu$ distribution to distinguish it from the standard CMP distribution. 


\begin{figure}
\label{fig:CMP-mu-distributions}
\includegraphics[width = \textwidth, trim={30mm 0 30mm 0}, clip]{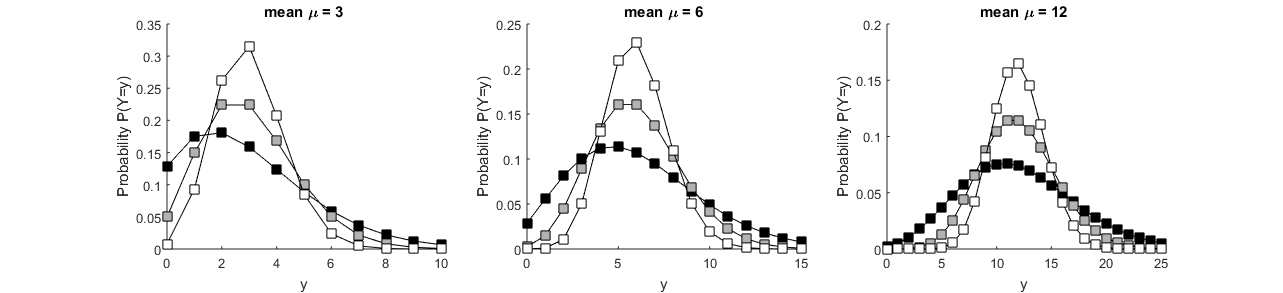}
\caption{Probability mass functions for underdispersed (white, $\nu = 2.1$), Poisson (grey, $\nu=1$), and overdispersed (black, $\nu =0.4$) CMP$_\mu$ distributions, with means $\mu = 3, 6$ and $12$.}
\end{figure}

The probability mass functions of CMP$_\mu$ distributions with various levels of dispersion and means are plotted in Figure 1. We see that a large dispersion $\nu$ condenses the distribution around the mean while a small dispersion stretches the distribution away from the mean, in a somewhat analogous manner to the precision parameter $1/\sigma^2$ in a Normal $N(\mu,\sigma^2)$ family. In fact, the mean $\mu$ and dispersion $\nu$ turn out to be orthogonal (see Result \ref{Result2}), just like the mean $\mu$ and scale $\sigma$ parameters in Normal families.

All the key features of CMP distributions hold for CMP$_\mu$ distributions also. For example, CMP$_\mu$ distributions generalize not only the Poisson distribution, but also other well-known discrete distributions. More precisely, CMP$_\mu$ distributions form a continuous bridge between the geometric, Poisson and Bernoulli distributions:
\begin{Result} 
\label{Result1}
As $\nu$ increases from 0 to $\infty$, CMP$_\mu$ models pass through the following special cases:
\vspace{-5mm}
\begin{center} 
\begin{tabular}{lcl}
$\nu =0$ & $\Longrightarrow$ & Geometric\{$p=1/(\mu+1)$\} \\
$\nu=1$ & $\Longrightarrow$ & Poisson\{$\mu$\} \\
$\nu \to \infty$ & $\Longrightarrow$ & Bernoulli\{$p = \mu$\}, provided $\mu \le 1$.
\end{tabular}
\end{center}
\end{Result}
Result \ref{Result1} can be established via direct calculations (see Appendix {\color{red}A}\ref{app:Result1} for details).

In general, $\nu <1$ implies overdispersion and $\nu > 1$ implies underdispersion relative to a Poisson distribution with the same mean. Note that $\nu=1$ is in the interior of the parameter space, so we can test whether a Poisson model is adequate using a standard likelihood ratio test (see Result \ref{Result7} below). In contrast, negative-binomial models (and other Poisson scale mixtures) cover the Poisson distribution only as a limiting case. While the $\nu = \infty$ case degenerates to a Bernoulli distribution, note that any finite dispersion $\nu < \infty$ corresponds to the support of $Y$ being the nonnegative integers $\mathbb{N}$. This is in contrast to the generalized Poisson distribution \citep{Famoye1993} which achieves underdispersion by a somewhat unnatural truncation of the support -- see \citet[][Section 1]{SS2010} for more discussion on this.

CMP$_\mu$ distributions can be shown to form two-parameter exponential families (see Appendix {\color{red}A}\ref{app:expfam}). This is analogous to the result from \citet[][Section 2.1]{SMKBB2005} for standard CMP distributions. For fixed $\nu$, CMP$_\mu$ distributions also form one-parameter exponential families (see Appendix {\color{red}A}\ref{app:expfam}), and it is this key property 
that makes them immediately adaptable to regression modelling via the generalized linear model \citep[GLM,][]{MN1989} framework .

In addition to retaining all the key features of standard CMP distributions, CMP$_\mu$ distributions also possess an attractive orthogonality property. This is particularly useful in regression settings, ensuring that the maximum likelihood estimator of the regression parameters will be asymptotically efficient and asymptotically independent of the estimated dispersion parameter (c.f. Results \ref{Result5} and {\color{red}5b}). Result \ref{Result2} below follows from Corollary 1 of \citet{HR2013}.

\begin{Result} [Orthogonality]
\label{Result2}
The mean $\mu$ and dispersion parameter $\nu$ in CMP$_\mu$ distributions are orthogonal.
\end{Result}

Finally, the current approach is also distinct from the reparametrization due to \citet{GC2008} which treats $\lambda^{1/\nu}$ as the location parameter. There is still no closed form for the mean under this latter parametrization, making it incompatible with simple, easily-interpretable mean-models, such as the log-linear model.

\section{CMP$_\mu$ regression models}
\vspace{-5mm}

Given a set of covariates $X \in \mathbb{R}^q$, a CMP$_\mu$ generalized linear model for count responses $Y$ can be specified via
\begin{equation}
\label{eq:compmureg}
Y \ |\ X \sim \mbox{CMP}_\mu(\mu(X^T\beta), \nu) \ ,
\end{equation}
where $\mu(.)$ is some mean-model and $\beta \in \mathbb{R}^q$ is a vector of regression coefficients. The examples in this paper focus mainly on the log-linear mean-model, 
$$E(Y|X) = \mu(X^T\beta) = \exp(X^T\beta)\ ,$$
although any other link function can be used, as is the case with any GLM. Note that model (\ref{eq:compmureg}) is a genuine GLM, so the score equations for the regression parameters have the usual ``weighted-least squares" form \citep[c.f.,][]{FK1985} and can be solved using standard Newton-Raphson or Fisher scoring algorithms (see Result \ref{Result4}). Moreover, all the familiar key features of GLMs \citep[e.g.,][Chapter 2]{MN1989} are retained.

The mean-dispersion specification in (\ref{eq:compmureg}) makes it directly comparable and compatible with a host of widely used log-linear regression models for counts. In particular, the mean $\mu = \exp(X^T\beta)$ is functionally independent of the dispersion parameter $\nu$, making it similar in structure to the familiar negative-binomial regression model for overdispersed counts. The dispersion $\nu$ can in turn be modeled via its own regression model using a set of covariates $\tilde X$, with the natural link being $\nu = \exp(\tilde X^T \gamma)$ to ensure non-negativity. The set of covariates $\tilde X$ can coincide with $X$, share common components with $X$, or be distinct altogether. 

A referee pointed out that the mean specification in (\ref{eq:compmureg}) also makes it easy to incorporate offsets into the model. For example, suppose the count responses are collected over spatial regions of possibly different areas or time periods of possibly different durations. The size of each unit of observation is generically called the ``exposure", and we can take into account different exposures by simply including offsets in the model,
$$
Y \ | \ X \sim \mbox{CMP}_\mu( \mbox{exposure} \times \mu(X^T\beta), \nu) \ ,
$$
much in the same way as for classical GLMs \citep[e.g.,][Page 206]{MN1989}. In contrast, handling offsets in standard CMP regression models is not as easy, precisely because standard CMP distributions are not parametrized via the mean. 

\section{Estimation and inference}
\label{se:theory}
\vspace{-5mm}

The log-likelihood for a generic observation $Y \sim$ CMP$_\mu(\mu, \nu)$ is
\begin{equation}
\label{eq:loglike}
l(\mu, \nu \ |\ Y) = Y \log(\lambda(\mu,\nu)) - \nu \log(Y!) - \log Z(\lambda(\mu,\nu), \nu) \ ,
\end{equation}
from which we can derive the score function for $\mu$ as
$$
S_\mu = \frac{\partial l}{\partial \mu} = \frac{\partial l}{\partial \lambda} \frac{\partial \lambda}{\partial \mu} = \frac{Y-\mu}{V(\mu, \nu)} \ ,
$$
where  
$$V(\mu, \nu) 
= \sum_{y=0}^\infty \frac{(y-\mu)^2 \lambda(\mu, \nu)^y}{(y!)^\nu Z(\lambda(\mu, \nu), \nu)}
$$
is the variance of $Y$; details of this derivation are given in Appendix {\color{red}A}\ref{app:score}. Thus, for independent and identically distributed (iid) data, $Y_1, Y_2, \ldots, Y_n$, the MLE for $\mu$ satisfies
$$
0 = \sum_{i=1}^n \frac{Y_i - \mu}{V(\mu, \nu)} \ ,
$$ 
so that the MLE is given simply by the sample mean,
$
\hat \mu = \bar{Y} \ .
$
In contrast, estimation of the rate parameter $\lambda$ in the standard CMP model is far more complicated, with \citet{SMKBB2005} giving three alternative estimation schemes for $\lambda$ for the iid case alone. Relatively simple estimation procedures for mean parameter(s) is a key feature of the CMP$_\mu$ parametrization.

\begin{Result}[iid case]
\label{Result3}
For iid data $Y_1,Y_2,\ldots, Y_n$, the MLE for $\mu$ is $\hat \mu = \bar{Y}$, the sample mean.
\end{Result}

For regression models where $\mu = \mu(X^T\beta)$ for some mean-model $\mu(.)$, covariate vector $X \in \mathbb{R}^q$ and parameter vector $\beta \in \mathbb{R}^q$, the score functions for the regression coefficients $\beta$
take on the usual weighted least-squares form \citep[c.f.,][]{FK1985},
$$
S_\beta = \frac{\partial l}{\partial \mu} \frac{\partial \mu}{\partial \beta} 
= \frac{Y - \mu(X^T\beta)}{V\left(\mu(X^T\beta), \nu\right)} \mu'(X^T\beta) X \ .
$$
We have the following result.
\begin{Result}[Regression case] 
\label{Result4}For independent data pairs $(X_1,Y_1), (X_2, Y_2,),\ldots, (X_n, Y_n)$, the MLE of $\beta$ can be characterized as the solution to the usual GLM score equations,
$$
0 = \sum_{i=1}^n  \frac{Y_i - \mu(X_i^T\beta)}{V\left(\mu(X_i^T\beta), \nu_i \right)} \mu'(X_i^T\beta) X_i \ .
$$
\end{Result}
As is the case with negative-binomial regression, the default is to assume a common dispersion $\nu$ across all data pairs. However, regression models for the dispersion are also possible \citep[see also][]{Smyth1989}. Corresponding score functions for the dispersion parameters are given in Appendix {\color{red}A}\ref{app:scorenu}.

The orthogonality property from Result \ref{Result2} implies that the MLE for the mean parameters $\beta$ are asymptotically efficient and asymptotically independent of the MLE for the dispersion parameters. As CMP$_\mu$ regression models are bona fide GLMs, we also have consistency and asymptotic normality of the MLE \citep[][]{FK1985}. We summarise in Result \ref{Result5} below.

\begin{Result}[Asymptotic normality, constant dispersion]
\label{Result5}
As $n \to \infty$,
$$
\sqrt{n} \left( \begin{array}{c}\hat \beta - \beta \\
\hat \nu - \nu 
\end{array} 
\right)
\stackrel{D}{\to} N \left( 0, \, \left[\begin{array}{cc} W_1 & 0 \\ 0 & W_2 
\end{array} 
\right] \right)\ ,
$$
where the asymptotic variance of $\hat \beta$ is
$$
W_1 = \left[ E^X \left(\frac{\mu'(X^T \beta)^2 X X^T}{V \left(\mu(X^T\beta), \nu \right)} \right) \right]^{-1} \ , 
$$
and the asymptotic variance $W_2$ of $\hat \nu$ is given in the Appendix.
\end{Result}

In Result 5, $E^X$ denotes expectation under the design measure for the covariates $X$. This can be either a prescribed design sequence or random sampling. Note that the asymptotic variance of $\hat \beta$ is the same as if the true dispersion $\nu$ were known to begin with, confirming $\hat \beta$ to be asymptotically efficient. Note also that $\hat \beta$ and $\hat \nu$ are asymptotically independent. A generalization of Result 5 for when the dispersion $\nu$ is itself modelled via a regression $\nu = \nu(\tilde X^T\gamma)$ is given in the Appendix.

MATLAB software for fitting CMP$_\mu$ regression models (\ref{eq:compmureg}) to data can be downloaded from the Online Supplement. The usual GLM form of the score equations (Result 4) means that standard Newton-Raphson algorithms can be used. This leads to substantial gains in computational efficiency, with the MATLAB routine being up to an order of magnitude faster than the state-of-the-art \texttt{cmp} function in the \texttt{COMPoissonReg} R-package for fitting standard CMP models, and over two orders of magnitude faster than the \texttt{hp.fit} function for fitting hyper-Poisson models in R (see computer run times in Section 5). 

The software also outputs, if requested, estimated standard errors for $\hat \beta$, obtained from the plug-in estimator of variance,
\begin{equation}
\label{eq:varest}
\hat{\mbox{Var}}(\hat \beta) =  \left[ \sum_{i=1}^n  \frac{\mu'(X_i^T \hat \beta)^2 X_i X_i^T}{V \left(\mu(X_i^T\hat\beta), \hat \nu \right)}\right]^{-1} \ .
\end{equation}
The loglikelihood value achieved at the maximum can also be outputted.

The estimated variance (\ref{eq:varest}) can be used to carry out inferences based on Wald-tests. An alternative is to use likelihood ratio tests (LRTs), which do not require estimating the variance, yet enjoy asymptotic optimality properties. Likelihood ratio tests are also particularly useful for testing categorical predictors, where it is sensible to either include all levels of a factor or to exclude the factor altogether. These can be carried out using the MATLAB routine by comparing the maximum loglikelihood values achieved with and without model constraints. More precisely, we have the following result, which is analogous to the analysis-of-deviance.

\begin{Result}[LRT for composite hypotheses]  
\label{Result6}Under $H_0: M \beta = \delta$, where $M$ is a $r \times q$ matrix of full rank and $\delta$ is an $r \times 1$ vector,
$$
2\left[\sup_{\beta, \nu} l_n(\beta, \nu) - \sup_{M \beta =\delta, \nu} l_n(\beta, \nu) \right] \stackrel{D}{\to} \chi^2_r \ , \mbox{ as } n \to \infty\ .
$$
A finite-sample adjustment can be made by comparing to a $r F_{r, n-q}$ distribution.
\end{Result}

We can also use likelihood ratio tests to examine whether a simpler Poisson model ($\nu=1$) is adequate.

\begin{Result}[LRT for testing Poisson]  
\label{Result7}Under $H_0: \nu = 1$,
$$
2\left[\sup_{\beta, \nu} l_n(\beta, \nu) - \sup_{\beta} l_n(\beta, 1) \right] \stackrel{D}{\to} \chi^2_1 \ , \mbox{ as } n \to \infty\ .
$$
\end{Result}

Results 1--7 suggest that CMP$_\mu$ models offer the best of both worlds -- they inherit all the desirable features of CMP models, including the flexibility to handle both over and underdispersion, with the parsimony, interpretability and familiarity of classical GLMs.

For post-fitting model diagnostics, residual plots based on Pearson or deviance residuals can be visually unappealing as they exhibit artificial banding due to the discrete nature of count responses. A probability inverse transformation (PIT) for discrete distributions was developed by \citet{Smith1985} as a better diagnostic tool for count models. If the underlying distributional model is correct, then the PIT should resemble a random sample from a standard uniform distribution. Goodness-of-fit can then be assessed by plotting a histogram of the PIT, or via a quantile plot of the PIT against the uniform distribution. 

\begin{figure}
\label{fig:GoodnessOfFit}
\begin{subfigure}{.325\textwidth}
\includegraphics[width = \textwidth, trim={65mm 115mm 60mm 117mm}, clip ]{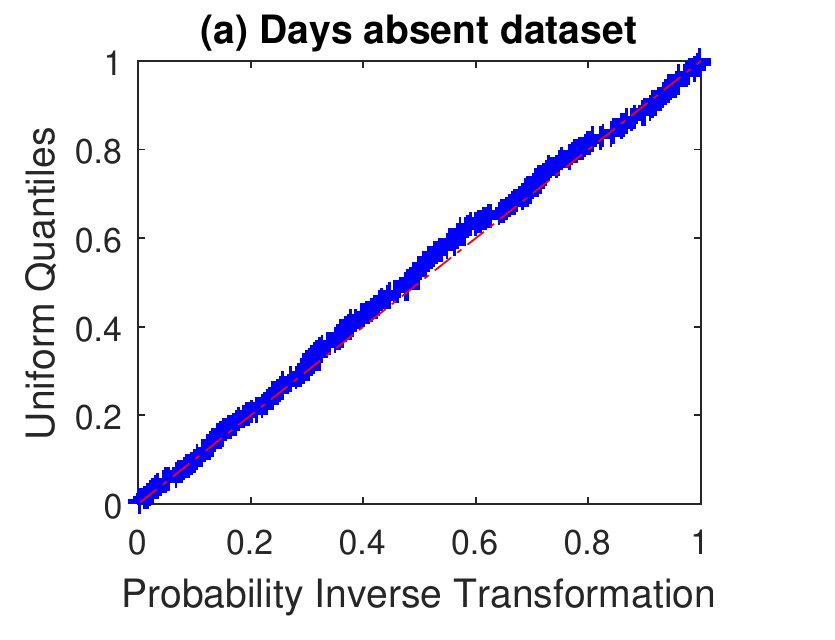}
\end{subfigure}
\begin{subfigure}{.325\textwidth}
\includegraphics[width = \textwidth, trim={65mm 115mm 60mm 117mm}, clip ]{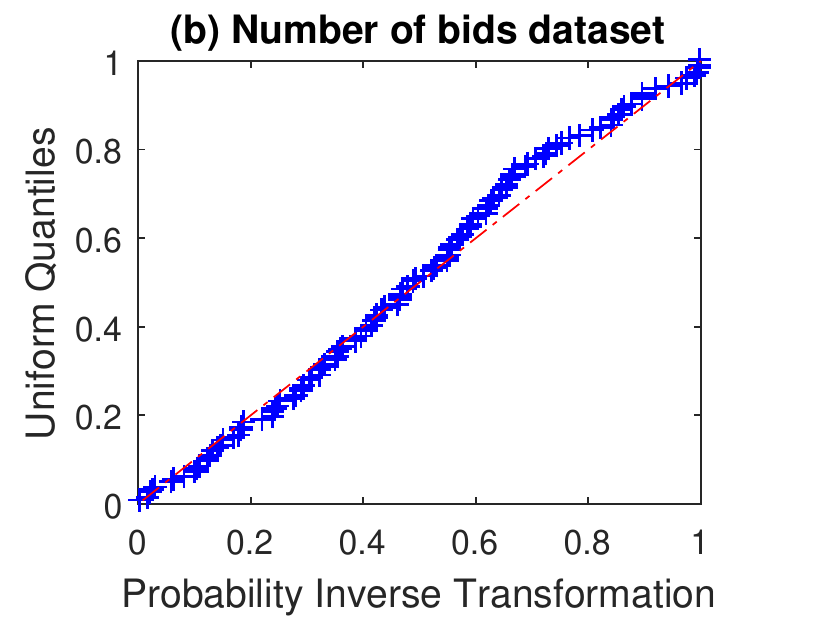}
\end{subfigure}
\begin{subfigure}{.325\textwidth}
\includegraphics[width = \textwidth, trim={65mm 115mm 60mm 117mm}, clip ]{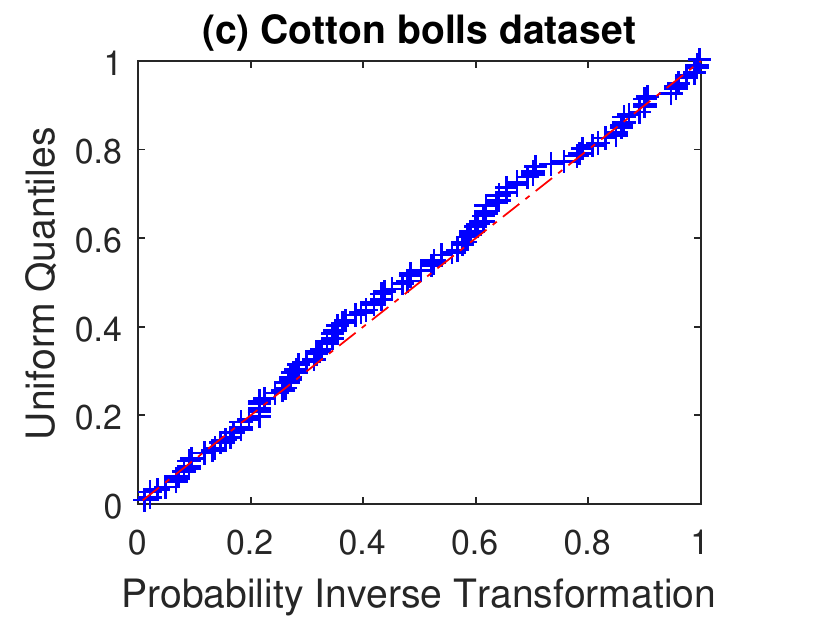}
\end{subfigure}
\caption{PIT-uniform quantile plots for fitted CMP$_\mu$ regression models for the (a) days absent, (b) takeover bids, and (c) cotton bolls count datasets from Sections \ref{se:daysabsdata}, \ref{se:numbidsdata} and \ref{se:cottonbollsdata}, respectively.}
\end{figure}

As a quick demonstration, three examples of PIT-uniform quantile plots are given in Figure 2. These come from three data analysis examples examined later in Section \ref{se:Examples}. In short, all three plots show reasonable closeness to uniformity, indicating that the fitted CMP$_\mu$ regression models are appropriate in each case. Detailed discussions of these three examples are given in Sections \ref{se:daysabsdata}, \ref{se:numbidsdata} and \ref{se:cottonbollsdata}, respectively. 

\section{Examples and simulations}
\label{se:Examples}
\vspace{-5mm}

We examine the finite-sample performance of the proposed method using various examples and a simulation study based on real datasets. We also compare to a range of competing models for dispersed counts, including the negative-binomial (Neg-Bin), generalized Poisson (GP), hyper-Poisson (hP), Gamma-count (GC) and standard CMP models. All models were fit using a Windows desktop with an i7-4770 CPU running at 3.40 GHz, and 16.0GB RAM. Neg-Bin models were fit using the \texttt{glm.nb} function in the \texttt{MASS} package in R. GP models were fit using the R code downloaded from Famoye's webpage (\url{http://people.cst.cmich.edu/famoy1kf/appliedstat/}). hP models were fit using the R function \texttt{hp.fit} downloaded from \url{http://www4.ujaen.es/~ajsaez/hp.fit.r} accompanying the paper by \citet{SS2013}. GC models were fit using the R function \texttt{pois2gc} from \url{http://www.leg.ufpr.br/~walmes/papercompanions/gammacount2014/papercomp.html} accompanying the paper by \citet{ZRBSM2014}. Standard CMP models were fit using the \texttt{cmp} function in the \texttt{COMPoissonReg} package in R. Finally, the CMP$_\mu$ model was fit using the MATLAB code provided in the Online Supplement. While some of these implementations are written on different platforms and perhaps to different levels of computational sophistication, they nevertheless represent the state-of-the-art for their corresponding methods. A comparison of these implementations is therefore of relevance and value to someone who is looking to use these methods today.

\subsection{Overdispersed class attendance data: analysis}
\label{se:daysabsdata}
\vspace{-5mm}

The attendance dataset from \url{http://www.ats.ucla.edu/stat/stata/dae/nb_data.dta} examines the relationship between the number of days absent from high school and the gender, maths score (standardized score out of 100) and academic program (``General", ``Academic" and ``Vocational") of 314 students sampled from two urban high schools. An insightful plot of the data, along with some summary statistics, can be found on the website \url{http://www.ats.ucla.edu/stat/r/dae/nbreg.htm}. These indicate that the number of days absent exhibits strong overdispersion. We first compare fits to the original data using log-linear negative-binomial, GP, CMP$_\mu$, hP, GC and standard CMP regression models. We then look at a simulation study based on the dataset. 

\begin{table}
\centering
\scriptsize
\caption{Attendance dataset -- estimated coefficients, standard errors (se), dispersion, AIC and computer run times for competing models. Sample size $n=314$.}
\begin{tabular}{l|rrrrrc}
& \multicolumn{1}{c}{\underline{Neg-Bin}} & \multicolumn{1}{c}{\underline{GP}} & \multicolumn{1}{c}{\underline{CMP$_\mu$}} & \multicolumn{1}{c}{\underline{CMP}} & \multicolumn{1}{c}{\underline{hP}} & \multicolumn{1}{c}{\underline{GC}} \\
Coefficient & estimate (se) & estimate (se)  & estimate (se) & estimate (se) & estimate (se) & estimate (se)  \\ \hline
intercept & 2.707 (0.204) & 2.673 (0.238) & 2.715 (0.190) & 0.018 (0.057) &  2.727 (0.180) \\
gender(Male) & -0.211 (0.122) & -0.194 (0.129) & -0.215 (0.117) & -0.035 (0.018) & -0.217 (0.116) & did \\
program(Academic) & -0.425 (0.182) & -0.421 (0.216) & -0.425 (0.170) & -0.050 (0.020) & -0.431 (0.161) & not \\
program(Vocational) & -1.253 (0.200) & -1.248 (0.228) & -1.254 (0.190) & -0.232 (0.041) & -1.261 (0.185) & converge \\
math & -0.006 (0.002) & -0.006 (0.003) & -0.006 (0.002) & -0.001 (0.000) & -0.006 (0.002) \\
\hline
dispersion & \multicolumn{1}{c}{1.047} &  \multicolumn{1}{c}{0.292} & \multicolumn{1}{c}{0.020} & \multicolumn{1}{c}{0.021} & \multicolumn{1}{c}{467.024}  \\
loglikelihood & \multicolumn{1}{c}{-864.2} & \multicolumn{1}{c}{-870.9}  & \multicolumn{1}{c}{-864.5} & \multicolumn{1}{c}{-864.0} & \multicolumn{1}{c}{-863.9}  \\
run time (seconds) & \multicolumn{1}{c}{0.05} & \multicolumn{1}{c}{0.07} & \multicolumn{1}{c}{6.7} & \multicolumn{1}{c}{41.4} & \multicolumn{1}{c}{3416}
\end{tabular}
\end{table}

The estimated coefficients, standard errors, dispersions, maximized loglikelihoods and computer run times from five of the six competing models (the GC model did not converge) are given in Table 1. We see that the negative-binomial, CMP$_\mu$, hP and standard CMP models all give similar maximized loglikelihood values, indicating comparable goodness-of-fits overall, with the GP model being somewhat inferior for these data. Moreover, the PIT quantile plot in Figure 2a shows close agreement with the uniform distribution, indicating that the CMP$_\mu$ model indeed fits the data well.

The negative-binomial, GP, CMP$_\mu$ and hP models are log-linear models, so they are easy to interpret. For example, the fitted CMP$_\mu$ model estimates that students in the General program are expected to miss $\exp(+1.254) = 3.5$ times more days of school compared to students in the Vocational program, with female students expected to miss $\exp(+0.215) = 1.2$ times more days of school compared to male students. A 10 point increase in maths scores is associated with a $\exp(-0.06) = 0.94$ times reduction in the expected days of absence from school. Finally, the estimated dispersion parameter of $\hat \nu = 0.020$ is much smaller than 1, reflecting the strong overdispersion exhibited by the data. Interpretations of the other log-linear models are analogous. While the hP model achieved the largest maximized loglikelihood here (and only slightly), it took over 3416 seconds, or 57 minutes, to fit. 

For the standard CMP model, the interpretation of coefficients is more opaque. The estimated dispersion parameter is $\hat \nu = 0.021 < 1$, so we can use the approximation
$$
\hat E(Y) \approx  \hat \lambda^{1/\hat \nu} - \frac{\hat \nu - 1}{2 \hat \nu}  = \hat \lambda^{47.6} + 23.31 \ ,
$$ where the rate $\hat \lambda$ is given by
\begin{eqnarray*}
\log(\hat \lambda) = 0.018 - 0.035 \, I(\mbox{Male}) - 0.050 \, I(\mbox{Academic}) - 0.232 \, I(\mbox{Vocational}) -  0.001 \, \mbox{math}  \ ,
\end{eqnarray*}
and $I(.)$ is the indicator function. We can ascertain from the fitted model the {\it direction} and {\it statistical significance} of the effect of each covariate -- for example, students with higher math scores tend to be absent significantly less often -- but we cannot give an interpretation of the coefficients in the form of mean contrasts.

The estimated dispersion under the CMP$_\mu$ and standard CMP regression models are essentially identical, indicating that both models are picking up similar levels of overdispersion despite modelling the counts on slightly different scales. This is consistent with Result \ref{Result2}, which states that the dispersion parameter is orthogonal to the mean model.

\subsection{Overdispersed class attendance data: simulations}
\vspace{-5mm}

We now assess the finite-sample accuracy of the competing methods using data simulated from the fitted negative-binomial model. More precisely, we first sample $n$ sets of covariates from the original dataset, then conditional on these covariates we generate count responses via
$$
Y \, | \, \mbox{gender, program, math} \sim \mbox{Neg-Bin} (\mu, \nu = 1.047) \ ,
$$ 
where the mean $\mu$ is given by the fitted model
\begin{eqnarray*}
\log(\mu) = 2.707 - 0.211 \, I(\mbox{Male}) - 0.425 \, I(\mbox{Academic}) 
- 1.253 \, I(\mbox{Vocational}) - 0.006 \, \mbox{math} \ .
\end{eqnarray*}
The sample sizes $n=50, 100$ and $314$ correspond to small, moderate, and the original sample sizes, respectively. 

For each synthetic dataset, we fit expanded models including all second-order interactions. The expanded model can be written in the model notation of \citet{MN1989} as
\begin{eqnarray*}
Y \sim \mbox{gender} + \mbox{program} + \mbox{math}  + \mbox{gender:program} + \mbox{gender:math} + \mbox{program:math} \ .
\end{eqnarray*}
We then examine Type I errors for simultaneously dropping all second-order interactions using the likelihood ratio test of Result 5, calibrated against the $F_{5,\, n-10}$ distribution. We compare these to Type I errors obtained from the negative-binomial, GP and standard CMP regression models, also calibrated against the $F_{5, \, n-10}$ distribution. We did not compare to the GC model as it failed to converge for the majority ($>$ 70\%) of our simulated datasets. We also did not compare to the hP model as it took over 10 minutes to fit each dataset of sample size $n=50$ and over 50 minutes for each dataset of sample size $n=314$. For the Neg-Bin, GP and CMP$_\mu$ models, a total of 5000 simulated datasets were used for each setting, but only 1000 simulations were used for the standard CMP model due to the slow computational speeds (see average run times in Table 2). Finally, note that the CMP$_\mu$, GP and standard CMP models are all misspecified for negative-binomial data.

\begin{table}
\centering
\scriptsize
\caption{Type I errors (\%) and average computer run times (seconds) for  dropping all second-order interactions using four competing methods on negative-binomial data. Sample sizes of $n=50, 100$ and $314$. $N=5000$ simulations per setting for Neg-Bin, GP and CMP$_\mu$ models. $N=1000$ simulations per setting for CMP model.}
\centering
\begin{tabular}{l|rrrcrrrcrrrcccc}
nominal level & \multicolumn{3}{c}{\underline{\hspace{7mm} $1\%$ \hspace{7mm} }} & & \multicolumn{3}{c}{\underline{\hspace{7mm} $5\%$ \hspace{7mm}}} & & \multicolumn{3}{c}{\underline{\hspace{7mm} $10\%$ \hspace{7mm}}} & & \multicolumn{3}{c}{\underline{average run time (s)}} \\
sample size & 50 & 100 & 314 & & 50 & 100 & 314  & & 50 & 100 & 314 & & 50 & 100 & 314\\
\hline
Neg-Bin & 1.5 & 1.4 & 1.0 & & 8.2 & 6.5 & 5.1 & & 16.4 & 12.0 & 10.5 & & 0.02 & 0.03 & 0.04 \\
 GP & 2.4 & 1.3 & 0.6 & & 9.2 & 6.2 & 3.1 & & 17.1 & 11.8 & 7.2 & & 0.03 & 0.05 & 0.13 \\
 CMP$_\mu$ & 1.5 & 1.5 & 1.4 & & 7.9 & 6.7 & 6.1 & & 15.5 & 12.2 & 12.3 & & 1.6 & 2.8 & 7.9 \\
 CMP & 3.0 & 1.8 & 2.8 & & 14.3 & 7.8 & 8.8 & & 20.1 & 15.3 & 16.6 & & 43.2 & 80.5 & 229.8  
\end{tabular}
\end{table}

The Type I errors at nominal 1\%, 5\% and 10\% levels for each method are displayed in Table 2. We first see that while the LRT based on the correctly-specified negative-binomial model is asymptotically exact, it can still be rather biased for small sample sizes. Of the misspecified models, the proposed CMP$_\mu$ model is the most robust, showing closest agreement with the negative-binomial Type I errors. For smaller sample sizes, the GP model does reasonably well but its Type I errors become rather conservative for large samples. The standard CMP model shows worst performance overall, reflecting the limitation of having to model the mean on a different scale.

Also displayed in Table 2 are the average computer run times for each simulation under each model. We see that the two CMP-based models are the most computationally expensive, but this is not unexpected given the flexibility these models offer. More interesting, perhaps, is that the proposed CMP$_\mu$ model is an order of magnitude faster to fit than the standard CMP model, which, when coupled with its increased parsimony and interpretability, makes it arguably a more appealing model to use in practice. This reinforces the earlier claim that CMP$_\mu$ models combine the attractive features of standard CMP models with the familiarity and simplicity of a GLM. The Neg-Bin and GP models were the fastest to fit, as these do not involve approximating a normalizing constant at each iteration.

\subsection{Underdispersed takeover bids}
\label{se:numbidsdata}
\vspace{-5mm}

A dataset from \citet{CJ1997} gives the number of bids received by 126 US firms that were successful targets of tender offers during the period 1978-85, along with the following set of explanatory variables \citep[descriptions taken from][]{SS2013}:
\begin{itemize}[noitemsep]
\item Defensive actions taken by management of target firm: indicator variable for legal defense by lawsuit (\texttt{leglrest}), proposed changes in asset structure (\texttt{rearest}),  proposed change in ownership structure (\texttt{finrest}) and management invitation for friendly third-party bid (\texttt{whtknght}).
\item Firm-specific characteristics: bid price divided by price 14 working days before bid (\texttt{bidprem}), percentage of stock held
by institutions (\texttt{insthold}), total book value of assets in billions of dollars (\texttt{size}) and book value squared (\texttt{size$^2$}).
\item Intervention by federal regulators: an indicator variable for Department of Justice intervention (\texttt{regulatn}).
\end{itemize}
Some summary statistics of the data are given in Table 6 in the Appendix. A key feature of the dataset is that it exhibits strong underdispersion after accounting for the explanatory variables. \citet{SS2013} used these data to show that their proposed hP regression model fits better than competing Poisson and order-1 Poisson polynomial models. Here, we compare fits to the data using the log-linear hP, GP, CMP$_\mu$ and standard CMP regression models. Interestingly, the GC model did not converge for these data either. The data are available from the \texttt{Edcat} R-package \citep{Croissant2011}.

The estimated coefficients, absolute $t$ statistics, dispersion parameters, loglikelihood values and computer run times from the four competing models are given in Table 3. We see that the hP model gives the best fit to these data in terms of the maximized loglikelihood, but this is not surprising as the example was chosen by \citet{SS2013} to highlight the hP method. While the CMP$_\mu$ model offers a slightly inferior fit to these data, it is better than the GP and standard CMP models. The PIT quantile plot in Figure 2b also shows good agreement with the uniform distribution, indicating that the CMP$_\mu$ model is indeed a good fit.

The hP model took 1481 seconds, or 24.7 minutes, to fit whereas the CMP$_\mu$ model took only 0.8 seconds. The standard CMP model took over 25 seconds to converge and offers an almost identical fit to the CMP$_\mu$ model. Of all these methods, the GP model was the fastest to fit, but its maximized loglikelihood was quite a bit lower. Note that the estimated dispersions from the CMP$_\mu$ and standard CMP models are again similar, reflecting the orthogonality property from Result \ref{Result2}.

\begin{table}
\centering
\scriptsize
\caption{Takeover bids dataset -- estimated coefficients, absolute $t$ statistics ($|t|$), dispersion, maximized loglikelihood and computer run times for competing models. Sample size $n=126$.}
\begin{tabular}{l|rrrrc}
& \multicolumn{1}{c}{\underline{hP}} & \multicolumn{1}{c}{\underline{GP}} & \multicolumn{1}{c}{\underline{CMP$_\mu$}} & \multicolumn{1}{c}{\underline{CMP}} & \multicolumn{1}{c}{\underline{GC}} \\
Coefficient & estimate \ ($|t|$) & estimate ($|t|$)  & estimate ($|t|$) & estimate ($|t|$)  & estimate ($|t|$) \\ \hline
intercept & 1.033 \  (2.61) &  0.939 \ (1.83) & 0.990 \ (2.27) & 1.836 \ (2.55) &  \\
leglrest & 0.246 \ (2.18) &  0.265 \ (1.82) & 0.268 \ (2.18) & 0.389 \ (2.04) &  \\
rearest & -0.276 \ (1.87) & -0.178 (0.97)  & -0.173 \ (1.12) & -0.297 \ (1.24) & did \\
finrest & 0.107 \ (0.65) & 0.067 \ (0.33) & 0.068 \ (0.39) & 0.112 \ (0.42) & not \\
whtknght & 0.492 \  (4.41) &  0.483 \ (3.14) & 0.481 \ (3.66) & 0.707 \ (3.38) & converge \\
bidprem & -0.704 \ (2.51) &  -0.646 \ (1.79) & -0.685 \ (2.23) & -1.015 \ (2.13) \\
insthold & -0.390 \ (1.17) &  -0.375 \ (0.92) & -0.368 \ (1.06) & -0.532 \ (1.02) \\
size & 0.179 \ (4.03) &  0.182 \ (3.27) &  0.179 \ (3.77) & 0.275 \ (3.30) \\
size$^2$ & -0.008 \ (3.25) & -0.008 \ (2.75) & -0.008 \ (3.05) & -0.012 \ (2.79) \\
regulatn & -0.010 \ (0.07) & -0.030 \ (0.20) & -0.038 \ (0.29) & -0.039 \ (0.20) \\
\hline
dispersion & \multicolumn{1}{c}{-2.624} &  \multicolumn{1}{c}{-0.021} & \multicolumn{1}{c}{1.754} & \multicolumn{1}{c}{1.727} \\
loglikelihood & \multicolumn{1}{c}{-170.2} & \multicolumn{1}{c}{-184.6}  & \multicolumn{1}{c}{-180.1} & \multicolumn{1}{c}{-180.4} \\
run time (seconds) & \multicolumn{1}{c}{1481} & \multicolumn{1}{c}{0.08} & \multicolumn{1}{c}{0.8} & \multicolumn{1}{c}{25.7}
\end{tabular}
\end{table}

\subsection{Underdispersed cotton boll counts}
\label{se:cottonbollsdata}
\vspace{-5mm}

\citet{ZRBSM2014} describe a dataset from a greenhouse experiment using cotton plants to examine the effect so five defoliation (def) levels (0\%, 25\%, 50\% 75\% and 100\%) on the observed number of cotton bolls produced on by the plants at five growth stages: vegetative, flower-bud, blossom, fig and cotton boll. The data exhibit strong underdispersion, which can be identified from the within-group means and variances as the experimental design was replicated five times (see Figure 3). Another plot of the data showing this underdispersion is given in Figure 2 of \citet{ZRBSM2014} . This motivated \citet{ZRBSM2014}  to formulate a Gamma-count (GC) process, which generalizes the Poisson process by replacing the exponential waiting times between successive counts with Gamma-distributed waiting times.

Here, we compare goodness-of-fits of the GC model to the log-linear GP, hP, CMP$_\mu$ and standard CMP regression models over a set of five nested linear predictors from \citet{ZRBSM2014} . In the following , the index $j \in \{\mbox{vegetative, flower-bud, blossom, fig, cotton boll}\}$ indicates the growth stage.
$$
\begin{array}{lccl}
\mbox{predictor I}: & g(\mu) &=& \gamma_0 ; \\
\mbox{predictor II}: & g(\mu) &=& \gamma_0 + \gamma_1 \mbox{def} ; \\
\mbox{predictor III}: & g(\mu) &=& \gamma_0 + \gamma_1 \mbox{def} + \gamma_2 \mbox{def}^2; \\
\mbox{predictor IV}: & g(\mu) &=&  \gamma_0 + \gamma_{1j} \mbox{def} + \gamma_2 \mbox{def}^2 \\
\mbox{predictor V}: & g(\mu) &=&\gamma_0 + \gamma_{1j} \mbox{def} + \gamma_{2j} \mbox{def}^2 .
\end{array}
$$
The link function $g$ is the $\log$ for the GC, GP, hP and CMP$_\mu$ models, but it has no closed form for the standard CMP model. Note that the parameter $\mu$ in the GC model characterizes the mean waiting time between successive counts rather than the mean of the count process. 

\begin{table}
\centering
\scriptsize
\caption{Cotton bolls dataset -- AIC and computer run times for five sets of predictors and five competing methods. Sample size $n=125$.}
\begin{tabular}{lcccccccrrrrr}
 & & \multicolumn{5}{c}{\underline{\hspace{20mm} AIC \hspace{20mm}}} & & \multicolumn{5}{c}{\underline{\hspace{10mm}  run time (seconds) \hspace{10mm} }} \\
predictor & np & GC & GP & hP & CMP$_\mu$ & CMP & & GC & GP & hP & CMP$_\mu$ & CMP  \\
\hline
I & 2 & 548.79 & 545.22 & 558.12 & 548.96 & & & 0.02 & 0.02 & 148 & 1.4 \\
II & 3 & 520.70 & 514.92 & 541.84 & 520.96 & 520.92  & & 0.03 & 0.02 & 213 & 1.6 & 5.5 \\
III & 4 & 519.96 & 514.28 & 542.32 & 520.20 & 520.18 & & 0.05 & 0.02 & 327 & 1.5 & 6.9 \\
IV & 8 & 456.29 & 460.32 & 520.28 & 456.48 & 456.40 & & 0.11 & 0.02 & 2488  & 6.7 & 18.5 \\
V & 12 & 440.77 & 453.75 & 522.42 & 440.82 & 440.50 & & 0.29 & 0.03 & 2317 & 8.6 & 31.5 \\ 
\hline

\end{tabular}
\end{table}

The AIC and computer run times for the five predictor models and five competing methods are given in Table 4. We see that the GC, CMP$_\mu$ and standard CMP models all give rise to essentially the same AICs for all 5 predictor models, indicating similar goodness-of-fits overall, with the GP and hP models being noticeably inferior. The CMP$_\mu$ model is a log-linear model for the mean count, and so is easier to interpret than the standard CMP model or the GC model, with the latter being a log-linear model for the mean time between counts and not the mean count directly. Moreover, the computation speed were three times faster for the CMP$_\mu$ model than the standard CMP model, and over 100 times faster than the hP model. This again reinforces the claim that the CMP$_\mu$ parametrization combines the flexibility of standard CMP models with the familiarity and computational efficiency of GLMs. Interestingly, the \texttt{cmp} software for standard CMP regression could not fit intercept-only models, which is why the entries for predictor I are empty.

\begin{figure}
\includegraphics[width=\textwidth]{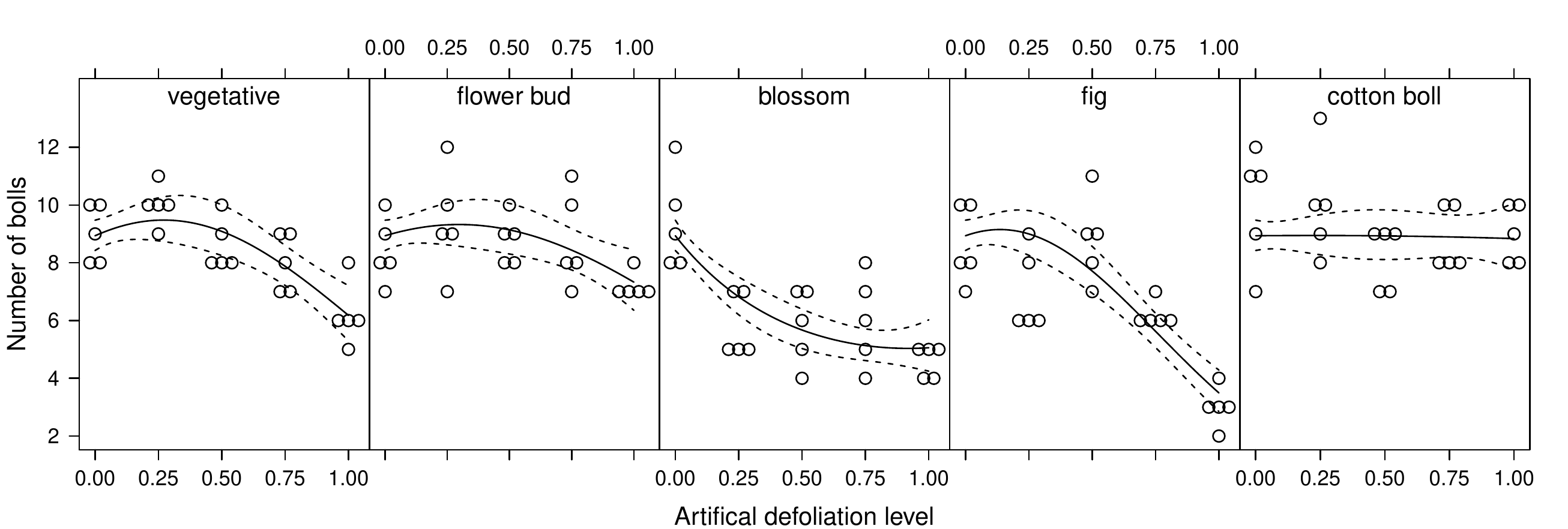}
\caption{Cotton bolls dataset -- observed values, fitted mean curves and 95\% confidence intervals for the CMP$_\mu$ model.}
\end{figure}

A plot of the predicted mean curves from the fitted CMP$_\mu$ model, along with corresponding 95\% confidence intervals, is given in Figure 2. These are essentially identical to those from the GC model \citep[c.f. Figure 3 in][]{ZRBSM2014}, indicating that the CMP$_\mu$ model matches the ``tailor-made" Gamma-count method for these data, in terms of both overall goodness-of-fit and subsequent inferences. This again demonstrates the flexibility and robustness of CMP$_\mu$ models for real data analysis problems. 

\begin{table}
\centering
\scriptsize
\caption{Cotton bolls dataset -- estimated coefficients, absolute $t$ statistics ($|t|$), dispersion parameters, maximized loglikelihood and computer run times for full model (predictor V) using five competing methods. Sample size $n=125$.}
\begin{tabular}{l|rrrrc}
& \multicolumn{1}{c}{\underline{GC}} & \multicolumn{1}{c}{\underline{GP}} & \multicolumn{1}{c}{\underline{CMP$_\mu$}} & \multicolumn{1}{c}{\underline{CMP}} & \multicolumn{1}{c}{\underline{hP}} \\
Coefficient & estimate \ ($|t|$) & estimate ($|t|$)  & estimate ($|t|$) & estimate ($|t|$)  & estimate ($|t|$) \\ \hline
intercept & 2.234 \ (79.7) &  2.200 \ (74.2) & 2.190 \ (74.6)  &  10.895 \ (7.76) & 2.196 \ (36.9) 
\\
$\gamma_{1\mbox{vegetative}}$ & 0.412 \ (1.81) &  0.413 \ (1.66) & 0.436 \ (1.82) & 2.019 \ (1.77) & 0.617 \ (1.28) \\
$\gamma_{2\mbox{vegetative}}$ &  -0.763 \ (2.95) & -0.796 \ (2.64) & -0.805 \ (2.96) & -3.724 \ (2.78) & -1.020 \ (1.88) \\
$\gamma_{1\mbox{bud}}$ &  0.274 \ (1.22) &  0.240 \ (1.01) & 0.288 \ (1.22) & 1.343 \ (1.21) &  0.090 \ (0.17)\\
$\gamma_{2\mbox{bud}}$ & -0.464 \ (1.85) & -0.443  \ (1.61) & -0.486 \ (1.85) & -2.265 \ (1.81) & -0.268 \ (0.50) \\
$\gamma_{1\mbox{blossom}}$ & -1.182 \ (4.43) & -1.366  \ (3.66)  &  -1.249 \ (4.41) & -5.748 \ (3.88) & -1.202 \ (2.18) \\
$\gamma_{2\mbox{blossom}}$ &  0.645 \ (2.15) &   0.796  \ (1.87) & 0.680 \ ( 2.13) & 3.133 \ (2.08) & 0.621 \ (1.00) \\
$\gamma_{1\mbox{fig}}$ & 0.320 \ (1.28) &  0.078  \ (0.26) & 0.351 \ (1.33) &  1.596 \ (1.30) & 0.370 \ (0.71)\\
$\gamma_{2\mbox{fig}}$ &  -1.199 \ (4.04) & -0.938  \ (2.46) & -1.290 \ (4.08) &  -5.894 \ (3.66) & -1.347 \ (2.19)\\
$\gamma_{1\mbox{boll}}$ &  0.007 \ (0.03) & -0.023  \ (0.10)  & 0.009 \ (0.04) & 0.038 \ (0.03) & -0.245 \ (0.51) \\
$\gamma_{2\mbox{boll}}$ & -0.018 \ (0.08) &  0.002  \ (0.01)  & -0.020 \ (0.08) & -0.090 \ (0.08) & 0.257 \ (0.49) \\
\hline
dispersion & \multicolumn{1}{c}{5.112} &  \multicolumn{1}{c}{-0.06} & \multicolumn{1}{c}{4.86} & \multicolumn{1}{c}{4.88} & \multicolumn{1}{c}{0.029} \\
AIC & \multicolumn{1}{c}{440.77} & \multicolumn{1}{c}{453.75}  & \multicolumn{1}{c}{440.82} & \multicolumn{1}{c}{440.50} &\multicolumn{1}{c}{522.43} \\
run time (seconds) & \multicolumn{1}{c}{0.29} & \multicolumn{1}{c}{0.33} & \multicolumn{1}{c}{8.5} & \multicolumn{1}{c}{31.5} &\multicolumn{1}{c}{2317}
\end{tabular}
\end{table}

For completeness, Table 5 displays the estimated coefficients, absolute $t$ statistics ($|t|$), dispersion, maximized loglikelihood and computer run times for the full model (predictor V) using the five competing methods. We see that the $t$ statistics from the CMP$_\mu$ model are essentially identical to those from the GC model, indicating that the CMP$_\mu$ model is making as efficient use of the data as the ``tailor-made" GC approach. Finally,  the PIT quantile plot for the full model in Figure 2c confirms that the CMP$_\mu$ model is indeed a good fit to the data.

\vspace{-7mm}
\section{Conclusion} 
\vspace{-5mm}

CMP distributions have already proven to be flexible and robust models for dispersed counts. By parametrizing CMP distributions via the mean, we have placed CMP models on the same level of interpretability and parsimony as other popular, competing count models, while retaining all the key features of CMP distributions that have made them increasingly attractive for the analysis of dispersed count data. We anticipate that the proposed CMP$_\mu$ parametrization will facilitate even wider adoption of CMP-based models in applied count regression problems, allowing them to be directly comparable to a host of competing log-linear models for counts. Moreover, the simple GLM structure of the proposed method makes it easy to understand on a conceptual level, and relatively easy to fit on a computational level, with computational times being significantly faster than for the similarly flexible hyper-Poisson and standard CMP models. While no one model outperformed every other model in the data analysis examples and simulation study considered in this paper, it should be noted that the CMP$_\mu$ model was always most comparable to the ``tailor-made" method for each of the examples. At the very least, CMP$_\mu$ count regression models should be a useful addition to any applied statisticians toolkit.

\vspace{-7mm}
\section*{Acknowledgements}
\vspace{-7mm}
The author thanks the Editor, Associate Editor and two anonymous for comments and suggestions that improved the paper. The author also thanks Dr Thomas Fung for help on probability inverse transforms for discrete distributions.

\appendix
\vspace{-5mm}
\section{Technical details and miscellaneous results}
\vspace{-5mm}

\subsection{Result \ref{Result1}}
\label{app:Result1}
\vspace{-5mm}

For the case $\nu = 1$, solving (\ref{eq:lambdamu}) gives $\lambda = \mu$ and the normalizing constant is $\sum_{y=0}^\infty \lambda^y/y! = e^\lambda = e^\mu$. Thus, $P(Y=y \, | \, \mu, 1) = e^{-\mu} \mu^y/ y!$ is the pmf of a Poisson distribution with mean $\mu$. 

For the case $\nu = 0$, solving (\ref{eq:lambdamu}) gives $\lambda = 1/(\mu + 1) < 1$ and the normalizing constant is $\sum_{y=0}^\infty \lambda^y = 1/(1-\lambda) = (\mu+1)/\mu$. Thus, $P(Y=y \, | \, \mu, 0) = \lambda^y(1-\lambda) = \mu/(\mu+1)^{y+1}$ is the pmf of a geometric distribution with probability parameter $p = 1/(\mu + 1)$. 

As $\nu \to \infty$, the term $(y!)^\nu$ in the denominator tends to $\infty$ for $y \neq 0, 1$, so that $Y$ only takes value $y=0$ or $y=1$ with probability proportional to $\lambda^0$ and $\lambda^1$, respectively. Thus, the CMP$_\mu$ distribution approaches a Bernoulli distribution with probability parameter $p = \lambda/(1+\lambda) = \mu$, from (\ref{eq:lambdamu}).

\subsection{CMP$_\mu$ distributions in exponential family form}
\label{app:expfam}
\vspace{-5mm}

Writing the pmf (\ref{eq:compmupdf}) as 
$$
P(Y=y \, | \, \mu, \nu) = \exp \big\{ y \log \left[\lambda(\mu,\nu) \right] - \nu \log\left(y! \right) - \log Z(\lambda(\mu,\nu), \nu) \big \} \ ,
$$
we see that CMP$_\mu$ distributions form two-parameter exponential families with sufficient statistics $Y$ and $\log(Y!)$ and corresponding natural parameters $\log \lambda(\mu, \nu)$ and $\nu$, respectively.  

For fixed dispersion $\nu$, we see that CMP$_\mu$ distributions also form one-parameter exponential families, with sufficient statistic $Y$ and natural parameter $
\log \lambda(\mu, \nu)$. Note that for a fixed or given dispersion $\nu$, the natural parameter and normalizing constant are viewed as functions of $\mu$ alone.

\subsection{Score function for $\mu$}
\label{app:score}
\vspace{-5mm}

First, recall that $Z = \sum_{y=0}^\infty \lambda^y/(y!)^\nu$, so that
$$
\frac{\partial Z}{\partial \lambda} = \sum_{y=0}^\infty y \frac{\lambda^{y-1}}{(y!)^\nu} = \frac{1}{\lambda} \sum_{y=0}^\infty y \frac{\lambda^{y}}{(y!)^\nu} = \frac{1}{\lambda} \mu Z \ ,
$$
by the definition of $\mu$ as the mean of the distribution. The derivative of the log-likelihood (\ref{eq:loglike}) with respect to $\lambda$ therefore simplifies to the expression
$$
\frac{\partial l} {\partial \lambda} = \frac{Y}{\lambda} - \frac{\partial Z/\partial \lambda}{Z} = \frac{Y - \mu}{\lambda} \ .
$$

Next, recall that $\lambda(\mu, \nu)$ is the solution to $0 = \sum_{y=0}^\infty (y-\mu) \lambda^y / (y!)^\nu$. Implicit differentiation with respect to $\mu$ gives
\begin{eqnarray*}
0 &=& - \left[ \sum_{y=0}^\infty \frac{\lambda^y}{(y!)^\nu} \right] + \left[ \sum_{y=0}^\infty (y-\mu) y \frac{\lambda^{y-1}}{(y!)^\nu}\right] \frac{\partial \lambda}{ \partial \mu} \\
&=& - Z + \left[ \frac{1}{\lambda} \sum_{y=0}^\infty (y-\mu) y \frac{\lambda^{y}}{(y!)^\nu}\right] \frac{\partial \lambda}{ \partial \mu} \\
&=& - Z + \left[ \frac{1}{\lambda} \sum_{y=0}^\infty (y-\mu)^2 \frac{\lambda^{y}}{(y!)^\nu}\right] \frac{\partial \lambda}{ \partial \mu} \ .
\end{eqnarray*}
The last equality is true because $\sum_{y=0}^\infty (y-\mu) \lambda^{y}/(y!)^\nu = 0$ by the definition of $\mu$ as the mean of the distribution. We therefore have $\partial \lambda/\partial \mu = \lambda/V(\mu, \nu)$, where 
$$V(\mu, \nu) = \sum_{y=0}^\infty \frac{(y-\mu)^2 \lambda(\mu, \nu)^y}{(y!)^\nu Z(\lambda(\mu,\nu), \nu)}
$$ 
is the variance of $Y$.

The score function for $\mu$ is therefore given by
$$
S_\mu = \frac{\partial l}{ \partial \mu} = \frac{\partial l}{\partial \lambda} \, \frac{\partial \lambda}{\partial \mu} = \frac{Y-\mu}{\lambda} \, \frac{\lambda}{V(\mu,\nu)} = \frac{Y-\mu}{V(\mu,\nu)} \ ,
$$
as claimed in Section \ref{se:theory}.
\subsection{Score function for $\nu$}
\label{app:scorenu}
\vspace{-5mm}

The score function for $\nu$ is
$$
S_\nu = \frac{\partial l}{\partial \nu} = E_{\mu,\nu}\big[\log(Y!) (Y-\mu) \big] \frac{(Y-\mu)}{V(\mu,\nu)} - \big[ \log(Y!) - E_{\mu,\nu}\log(Y!)\big] \ ,
$$
which is immediately seen to be unbiased, i.e., $E S_\nu = 0$. For notational convenience, define operators $A$ and $B$ by $A(\mu, \nu) = E_{\mu,\nu}[\log(Y!) (Y-\mu)]$ and $B(\mu, \nu) = E_{\mu,\nu}\log(Y!)$, respectively, so that $S_\nu$ can be expressed more succinctly as
$$
S_\nu = A(\mu, \nu) \frac{Y-\mu}{V(\mu, \nu)} - [\log(Y!) - B(\mu,\nu)] \ .
$$

\subsection{MLE for $\nu$ for iid data}
\vspace{-5mm}

For iid data $Y_1, Y_2, \ldots, Y_n \sim $ CMP$_\mu(\mu,\nu)$, the MLE $\hat \nu$ can be characterised as the solution to $\mathbb{P}_n S_\nu = 0$, where $\mathbb{P}_n = n^{-1} \sum_{i=1}^n$ is the empirical expectation operator. This can be simplified to a moment-matching equation,
$$
\frac{1}{n}\sum_{i=1}^n \log (Y_i !) \, = \, E_{\bar Y, \nu} \log(Y!) \, = \,  B(\bar{Y}, \nu) \ .
$$

\subsection{MLE for $\nu$ in the regression case, and asymptotic variance of $\hat \nu$ from Result \ref{Result5}}
\vspace{-5mm}

For regression problems with independent data pairs $(X_1, Y_1), (X_2, Y_2), \ldots, (X_n, Y_n)$, the MLE $\hat \nu$ is characterised as the solution to the score equation
$$
0 = \sum_{i=1}^n \left\{ A \left(\mu(X_i^T\hat\beta), \nu \right) \frac{Y_i-\mu(X_i ^T\hat\beta)}{V \left(\mu(X_i^T\hat\beta), \nu \right)} - \left[\log(Y_i!) - B \left(\mu(X_i^T\hat\beta),\nu \right)\right] \right\} \ ,
$$
and its asymptotic variance $W_2$ is given by
$$
W_2^{-1} = E^X \left[\frac{A \left(\mu(X^T\beta), \nu^* \right)}{V \left(\mu(X^T\beta),\nu\right)} + C\left(\mu(X^T\beta), \nu\right) \right] \ ,
$$
where $C(\mu, \nu) = \mbox{Var}_{\mu,\nu}(\log(Y!))$ and $E^X$ denotes expectation over the design measure of $X$.

\subsection{Generalization of Result \ref{Result5} for $\nu = \nu(\tilde X^T\gamma)$.}
\vspace{-5mm}

If $\nu$ itself is modelled via a regression $\nu = \nu(\tilde X^T\gamma)$ for some function $\nu(.)$, covariates $\tilde X \in \mathbb{R}^{\tilde q}$ and corresponding parameter vector $\gamma$, then the score function for $\gamma$ is given by
$$
S_\gamma = S_\nu \, \nu'(\tilde X^T \gamma) \tilde X \ .
$$
We then have the following generalization of Result \ref{Result5}.
\begin{Result5b}[Asymptotic normality, non-constant dispersion] As $n \to \infty$,
$$
\sqrt{n} \left( \begin{array}{c}\hat \beta - \beta \\
\hat \gamma - \gamma
\end{array} 
\right)
\stackrel{D}{\to} N \left( 0, \, \left[\begin{array}{cc} W_1 & 0 \\ 0 & W_2 
\end{array} 
\right] \right)\ ,
$$
where the asymptotic variance $W_1$ of $\hat \beta$ is given by
$$
W_1^{-1} = \left[ E^{X, \tilde X} \left(\frac{\mu'(X^T \beta)^2 X X^T}{V \left(\mu(X^T\beta), \nu(\tilde X^T \gamma) \right)} \right) \right] \ , 
$$
the asymptotic variance $W_2$ of $\hat \gamma$ is given by
$$
W_2^{-1} = E^{X, \tilde X} \left[ \left\{ \frac{A \left(\mu(X^T\beta), \nu(\tilde X^T\gamma) \right)}{V \left(\mu(X^T\beta),\nu(\tilde X^T\gamma) \right)} + C\left(\mu(X^T\beta), \nu(\tilde X^T\gamma)\right) \right\}  \nu'(\tilde X^T\gamma)^2 \tilde X \tilde X^T \right] \ ,
$$
and $E^{X,\tilde X}$ denotes the expectation with respect to the design measure of $(X, \tilde X)$.
\end{Result5b}

\subsection{Summary statistics for the takeover bids dataset}
\vspace{-2mm}

\begin{table}[h]
\centering 
\scriptsize
\caption{Takeover bids dataset -- summary statistics}
\hspace{15mm}
\adjustbox{valign=t}{\begin{minipage}{.4\linewidth}
\begin{tabular}{c|ccrc}
\multicolumn{5}{c}{\underline{\quad  Numerical variables \quad }}  \\
variable & min & max & mean & sd \\
\hline
numbids & 0 & 10 & 1.74 & 1.43 \\
bidprem & 0.94 & 2.07 & 1.35 & 0.19 \\
insthold & 0 & 0.90 & 0.25 & 0.19 \\
size & 0.02 & 22.17 & 1.22 & 3.10
\end{tabular}
\end{minipage}}
 \hspace{10mm}
\adjustbox{valign=t}{\begin{minipage}{.4\linewidth}
\begin{tabular}{l|c}
\multicolumn{2}{c}{\underline{\quad  Binary variables \quad }} \\
 variable & percentage \\
\hline
leglrest & 42.9\% \\
rearest & 18.3\% \\
finrest & 10.3\% \\
regulatn & 27.0\% \\
whtknght & 59.5\%
\end{tabular}
\end{minipage}}
\end{table}




\end{document}